\documentclass[journal=jacsat,manuscript=article]{achemso}

\usepackage{chemformula} 
\usepackage[T1]{fontenc} 

\author{Andrey K. Sarychev}
\email{sarychev_andrey@yahoo.com}
\affiliation{Institute for Theoretical and Applied Electrodynamics, 
Russian Academy of Sciences,125412, Moscow, Russia}
\author{Andrey V. Ivanov}
\email{av.ivanov@physics.msu.ru}
\affiliation{Institute for Theoretical and Applied Electrodynamics, 
Russian Academy of Sciences,125412, Moscow, Russia}
\author{Gr\'egory Barbillon}
\email{gregory.barbillon@epf.fr}
\affiliation{EPF-Ecole d’Ingénieurs, 92330 Sceaux, France}

\title[An \textsf{achemso} demo]
  {Limit possible electric field in plasmon nanogap resonator}


\begin{document}

\begin{abstract}
We propose the theory of the plasmon excited in 
ultra-narrow plasmonic
 gap formed by a metal cylinder on a metal surface, 
 that generates a huge resonant local electromagnetic 
 field in optical frequencies. Resonance conditions are found. 
 The maximum possible enhancement of
 the electric field in the nanogap is estimated as 
 $\left|E_{max} \right|/ \left| E_{0} \right| \propto 
 \left| \varepsilon_m \right|^{3} 
 \varepsilon_{d}^{-1} \left( \Im \varepsilon_{m}\right)
 ^{-2}$.
 A simple  ultranarrow-gaped resonator can be used 
 for SERS sensing and infrared spectroscopy 
 of adsorbed molecules. 
 We assume that a plasmonic nanogap 
 is simplest and most straightforward 
 way to increase flexibility as well as sensitivity of the 
 plasmon-enhanced spectroscopes. 
\end{abstract}

\section{Introduction}
Calculation of the electric field 
in the gap between two metallic
 cylinders is a classical problem 
 of the classical electrodynamics. 
 However, the metal permittivity is typically negative 
 in optical spectral range, and plasmons are excited 
 between the cylinders, 
 which give a big enhancement 
 of the local electric field in the crevice
\cite{Vidal1996}. 
Later, it was shown \cite{Gresillon1999} 
that giant optical field fluctuations 
in random semicontinuous metallic films 
are due to the local
 field concentration in-between large 
 metallic clusters. 
A precise separation of plasmonic 
nanostructures is of
 widespread interest, because it controls the build-up 
of intense and localized optical fields, as well as their 
spectral tuning \cite{Ciraci2012,Moreau2012}. 
Plasmonic enhancement can also be used to improve
the efficiency of photodetectors 
\cite{Echtermeyer2011}.
Similarly, the non-linear frequency conversion 
critically 
depends on the field enhancements, their spatial 
localization,
and their spectral resonances whose all 
ultra-sensitivity 
depends on the plasmonic gaps \cite{ZhangY2011}.
Fundamental processes like
quantum tunneling can be observed optically, 
but only for gap separations 
in the subnanometer regime \cite{Savage2012}. 
Graphene could be used as the thinnest  
spacer between
gold nanoparticles and a gold substrate 
\cite{MertensJ2013}. 
The atomic layer deposition of the $Al_2 O_3$ 
film provides a subnanometer layer that has been 
employed 
to successfully fabricate high quality metallic 
nanogap 
arrays with sub-5 nm gap size \cite{CaiH2016}. 
In the works 
\cite{MubeenS2012,ChikkaraddyR2017,%
ChikkaraddyR2018,DemetriadouA2017}, 
high local electric fields were generated by 
ultra-narrow gap that 
formed by gold nanoparticle placed on gold mirror. 
Evolution of  different gap modes with 
a gap size of 1-5\,\textit{nm} was 
considered. It was shown that the coupling of 
transverse and antenna modes is altered  by varying 
the gap size and the 
nanoparticle shape from sphere 
to cube, 
resulting in strongly hybridized 
modes. Theoretical predictions 
and experimental studies have 
shown giant electromagnetic field 
fluctuations in case of almost 
touched plasmonic nanoparticles 
\cite{Genov2004,SealK2005,IvanovA2012,%
Frumin2013,%
LiuZ2013,LiuG2014,Liu2014,Rasskazov2013}. 
It was shown that plasmonic nanocavities 
confining the light to 
unprecedentedly small volumes, 
support multiple types of modes. 
Different nature of these modes 
leads to mode beating within the 
nanocavity and the Rabi oscillations, which alters the 
spatio-temporal dynamics of the hybrid system 
\cite{DemetriadouA2017}. 
By intermixing plasmonic excitation in 
nanoparticle arrays with excitons in a 
$WS_2$ monolayer inside a 
resonant metallic microcavity, 
the hierarchical system was 
fabricated with the collective
 microcavity-plasmon-exciton Rabi 
splitting exceeding 
$0.5\, eV$ at room temperature. Gap-surface 
plasmon metasurfaces, which consist of a 
subwavelength thin 
dielectric spacer sandwiched between an optically 
thick film of 
metal and arrays of metallic 
subwavelength elements arranged in a 
strictly or quasiperiodic fashion, 
have a possibility to fully control 
the amplitude, phase, and polarization 
of the reflected light 
\cite{DingF2018}. 
Such systems are successfully used for 
surface-enhanced infrared absorption 
\cite{DongL2017}. Strong 
motivation for the investigation 
of the gap plasmons is the SERS effect,
which relies on the 
plasmonic enhancement to enable 
identification of trace molecules 
captured within  gaps. The SERS 
is extremely important 
for medical diagnostics, for instance, 
for cancer detection, imaging and therapy, 
drug delivery, 
quantitative control of biomarkers 
including glycated proteins 
and cardiovascular biomarkers 
\cite{Kneipp2017,HuY2017,AndreouC2016,%
NechaevaNL2020,ChonH2014}. 
In recent years, 
there are many computer simulations as well as experimental 
works that are devoted to 
perfect coupled particles with 
ultra-narrow gaps 
\cite{CaiH2016,%
FuQ2015,JiangT2018,LiuG2018,%
LuX2018,MaC2016,%
NamJM2016,PanR2018,%
ShinY2015,SigleDO2015,%
YooD2018,ZhouJ2016}. 
In this paper, we propose the
theory of the plasmons excited in a ultra-narrow 
gap formed by a metallic 
cylinder on a metallic surface for different metals. 
We find the resonance conditions, when
 it is possible to 
achieve possible limit electromagnetic 
field both inside and in 
the vicinity of the gap. 
This will increase the sensitivity 
of the SERS-probing as well as this of 
other existing surface enhanced spectroscopes.

\section{Analytical theory \label{Analytic}}%
\begin{figure}[ht!]
\begin{center}
\includegraphics[scale=0.37]{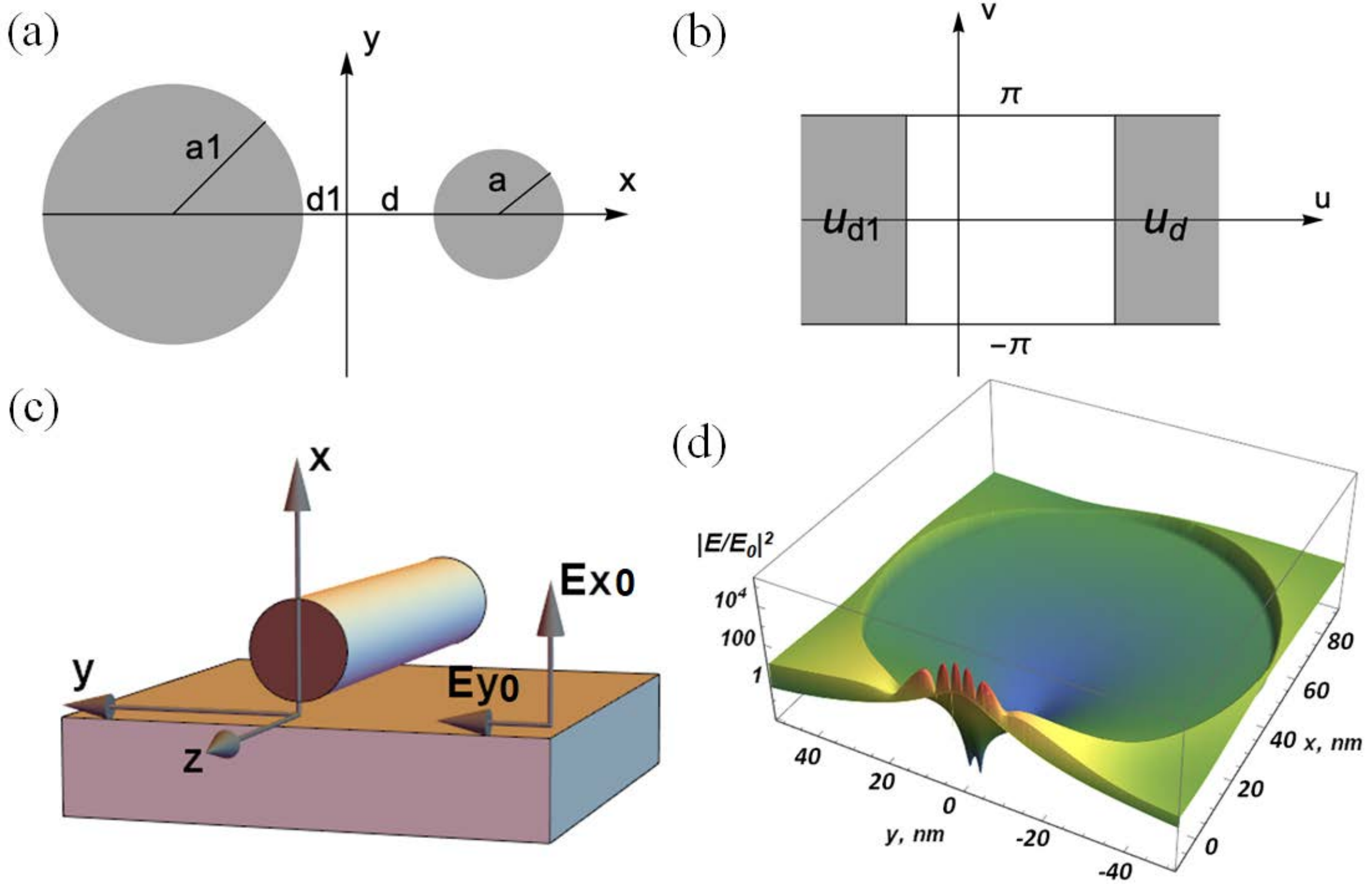}
\caption{The design of cylindrical 
nanoparticle-on-mirror cavity. (a-b) Conformal map 
of two cylinders into $u-v$ strip; (c) metallic 
cylinder above metal plate, electromagnetic wave 
is incident from the top; 
(d) electric field between 
Ag plate and cylinder with a radius of $a = 
50 \,nm$, gap $ d= 0.5\,nm $ calculated at
the excitation wavelength of
$\lambda = 405 \, nm$.
\label{Fig1}}
\end{center}
\end{figure}

We consider two parallel metallic 
cylinders whose respective 
radii are 
$a$
 and 
 $a_1 > a$
that are separated by the distance 
$d + d_{1 }$ 
so that the spacing between the cylinder axes is 
$a + d + a_1 + d_{1}$. 
The $x$-axis connects the cylinder centers so the 
centers have coordinates $\{x, y\} = \{a + d, 0\}$ and 
$\{-a_1 - d_1, 0\}$, correspondingly, as shown in 
Fig.1. To find the optical electric field in the nanogap 
between the cylinders illuminated by the incident 
light (see Fig.\ref{Fig1}d), we use new coordinates $u$ and $v$ that 
convenient to introduce in the complex form $z = L 
\tanh w/2$, where $z= x + i y$, $w = u + i v$, and the 
characteristic scale $L = \sqrt {d (2 a + d )}$, where 
$d$ is the distance from the surface of the cylinder 
``$a$" to the origin of the coordinates. Note $a$, 
$a_{1}$, and $L$ fully define the system geometry, if 
we define 
$d_{1} = \sqrt{a_{1}^{2} +L^{2} } -  a_{1} < d$. 
In the transition from 
$z$ to $w$, 
the whole plane $\{x,y\}$ transforms to the strip 
$-\infty < u < \infty$, $-\pi < v < \pi$. The surfaces of 
the cylinders  ``$a$" and  ``$a_1$" are transformed to 
the vertical lines $u_{d} = \ln [(a+d)/(a-d)] >0$ and 
$u_{d_{1}} = - \ln [(a_1 + d_1)/(a_1 -d_1)] <0$ 
(Fig.\,\ref{F0}).
All the space outer of the cylinders shrinks 
into the rectangular 
$[ u_{d_{1}} < u < u_{d},\; -\pi < v < \pi ]$. 
The approach is similar to the map used in 
\cite{IvanovA2012,KraftM2015}. 
The solution of the Laplace equation in the 
rectangular can be 
presented as sum of the functions 
$f_{Rc}^{(q)} = e^{-q u} \cos q v, \; 
 f_{Rs}^{(q)} = e^{-q u} \sin q v, \; 
 f_{Lc}^{(q)} = e^{q u} \cos q v, \; 
 f_{Ls}^{(q)} = e^{q u} \sin q v$, $q=1,2 \ldots$ :
\vspace{-10 pt} 
\begin{equation}\label{F0}
\varphi^G (u,v) = \varphi_0 + \\
\sum_{q = 1}^{\infty} \left[A_1^{(q)} 
f_{Rc}^{(q)}+A_2^{(q)} f_{Rs}^{(q)}+A_3^{(q)} 
f_{Lc}^{(q)}+A_4^{(q)} f_{Ls}^{(q)} \right]
\end{equation}
where $\varphi_0$ is the potential of the external electric field. We assume that the electric field concentrates in the nanogap whose the size is much smaller than the wavelength of the incident light $d+d_1 \ll \lambda$. Electric potentials in the  ``$a$" and  ``$a_1$" cylinders are given by the following expressions, respectively:
\vspace{-10 pt} 
\begin{equation}\label{F1}
\varphi^R (u,v) = \varphi_0 + \sum_{q = 1}^{\infty} 
\left[B_1^{(q)} f_{Rc}^{(q)}+B_2^{(q)} f_{Rs}^{(q)} 
\right],\;\; u > u_{d}
\end{equation}
\vspace{-20pt}
\begin{equation}\label{F2}
\varphi^L (u,v) = \varphi_0 + \sum_{q = 1}^{\infty} 
\left[B_3^{(q)} f_{Lc}^{(q)}+B_4^{(q)} f_{Ls}^{(q)} 
\right],\;\; u < u_{d_{1}}
\end{equation}
The components of the electric field in $\{ u, v\}$ 
space 
$\{ E_u, E_v \} = - \nabla_w  \varphi =
  -\{ \partial \varphi /\partial u,$ $\partial 
  \varphi /\partial v 
\}$ are complex since the metal permittivity 
is complex.
 Electric field in $\{ x, y\}$ space equals to 
$\{ E_x, E_y \} = \hat{J} \{ E_u, E_v \}$, where the 
jacobian $\hat{J}$ is obtained from the derivative 
$w_z = dw/dz$, namely, $J_{11} = J_{22} = \Re 
w_z$, $J_{21} =- J_{12} = \Im w_z$.
The  coefficients $A^{(q)}$ and $B^{(q)}$ in 
Eqs.\,(\ref{F0}), (\ref{F1}) and (\ref{F2}) are found 
from the boundary conditions 
for the electric fields 
$E^{R} = -\nabla_w  \varphi^{R}$, 
$E^{G} = -\nabla_w  \varphi^{G}$, 
and 
$E^{L} = -\nabla_w \varphi^{L}$
 at the surface of  ``$a$" and  ``$a_1$" cylinders:
\begin{equation}\label{RightBoundary}
E^{R}_{v} = E^{G}_v, \;
\varepsilon_m E^{R}_{u}  = 
\varepsilon_d E^{G}_{u}, \; u = u_d
\end{equation}
\vspace{-20pt}
\begin{equation}\label{LeftBoundary}
E^{L}_{v}  = E^{G}_v, \;
\varepsilon_{m_{1}} E^{L}_{u} = 
\varepsilon_d E^{G}_{u} ,
\; u = u_{d_{1}}
\end{equation}
where 
$\varepsilon_m$,  $\varepsilon_{d}$, 
and 
$\varepsilon_{m_{1}}$
are the permittivities of the 
 ``$a$"
cylinder, the outer space, and the ``$a_1$" cylinder, 
respectively. We expand the external field 
$E_0 = $ $ -\nabla \varphi_0 $
$ =\{ E_{0x},  E_{0y} \}$
 in a series of the functions 
 $f_{Rc}^{(q)}, \, f_{Rs}^{(q)}, $ 
 $f_{Lc}^{(q)} ,  \, f_{Ls}^{(q)}$. 
 For the simplest case of the constant external field,
the expansion 
$\left\{E_{0u}, E_{0v} \right\} = $
$  \sum_q  \left\{E_{0u}^{(q)}, 
 E_{0v}^{(q)} \right\}$, 
 has the following form:
\vspace{-5pt} 
\begin{equation}\label{E0u} 
\left\{E_{0u}^{(q)}, E_{0v}^{(q)} \right\} =
 -2 L (-1)^q q \times \\
\left\{\begin{matrix}
E_{0x} f_{Rc}^{(q)} - E_{0y} f_{Rs}^{(q)},\;
E_{0x} f_{Rs}^{(q)} + E_{0y} f_{Rc}^{(q)}, \; 
u > 0 \\
E_{0x} f_{Lc}^{(q)} + E_{0y}
 f_{Ls}^{(q)},\;E_{0y} f_{Lc}^{(q)}
  - E_{0x} f_{Ls}^{(q)}, \; u < 0
\end{matrix}  \right\}
\end{equation}
where 
$q = 1,2, \ldots$. 
We substitute the above equation 
in the boundary equation 
(\ref{RightBoundary}) for the surface 
of the  ``$a$" cylinder and equate the coefficients at the same $f^{(q)}$ functions obtaining the following equations for the coefficients $\{ A^{(q)}, B^{(q)} \}$ in Eqs.\,(\ref{F0}), (\ref{F1}) and (\ref{F2}):
\vspace{-5pt} 
\begin{multline}\label{AB}
\varepsilon_d \left(A_3^{(q)}+
A_1^{(q)} g ^{2 q} \right) - \varepsilon_m B_1^{(q)} 
= E_{0x} \left(\varepsilon_m-\varepsilon_d\right), \\
\varepsilon_d (A_4^{(q)} - A_2^{(q)} g ^{2 q})
 - \varepsilon_m B_2^{(q)} = 
 E_{0y} \left( \varepsilon_m-\varepsilon_d \right), \\
A_2^{(q)} g ^{2 q}+A_4^{(q)} - B_2^{(q)}=
A_1^{(q)} g ^{2 q} - A_3^{(q)} + B_1^{(q)}=0 
\end{multline}
where $g  = (L + d)/(L - d)$. Matching electric field 
and displacement at the surface of the ``$a_{1}$" 
cylinder, we obtain from Eq.\,(\ref{LeftBoundary}) :
\vspace{-10pt} 
\begin{multline}\label{AB1}
\varepsilon_d \left(A_1^{(q)}+
A_3^{(q)} g_1^{2 q}\right) -
 \varepsilon_{m_{1}} B_3^{(q)}  =
E_{0x} \left(\varepsilon_{m_{1}} 
-\varepsilon_d\right),\\ 
\varepsilon_d \left(A_2^{(q)}-A_4^{(q)}
 g_1^{2 q}\right) - \varepsilon_{m_{1}} B_4^{(q)}
  = E_{0y} \left(\varepsilon_{m_{1}}-
\varepsilon_d\right),\\
A_4^{(q)} g_1^{2 q}+A_2^{(q)}-B_4^{(q)}=
A_3^{(q)} g_1^{2 q} -A_1^{(q)}+B_3^{(q)}=0 
\end{multline}
where $g_1  =  (L + d_1)/(L - d_1)$. Zeroing of the 
determinant of Eqs.\,(\ref{AB}) and (\ref{AB1}) gives 
the condition for the  ``$q$" resonance. The cylinders 
with the same permittivity 
$\varepsilon_{m_{1}}  = \varepsilon_{m}$ 
resonate when 
$( g  g_{1} )^{2 q} \left( \varepsilon_d +
 \varepsilon_m \right)^{2} - 
 \left( \varepsilon_m - \varepsilon_d \right)^{2} = 0$. 
 The first term dominates when the distance 
 $d + d_1$ 
 between the cylinders increases and 
 $g  g_{1} 	\rightarrow \infty$.  
 Then, all the resonance frequencies 
 $\omega_{r}^{(q)}$ collapse to the single value 
 $\omega_{r}^{(1)}$ given by the well-known 
 equation 
 $\Re [\varepsilon_d (\omega_{r}^{(1)}) +
  \varepsilon_m (\omega_{r}^{(1)})] =0$ 
for the plasmon resonance in a metallic cylinder. 
On the other hand, when the gap size vanishes 
$d + d_1  \rightarrow 0$, the factor 
$g  g_{1} \rightarrow 1$ from above and the 
resonance frequencies 
$\omega_{r}^{(q)}$ spread 
out from $\omega_{r}^{(1)}$ to the minimum resonance
frequency estimated from the equation 
$\varepsilon_m (\omega_{r}^{(m)}) \simeq
-\frac{2 \varepsilon_{d}  a a_{1} }{ L(a + a_{1})} 
 \rightarrow - \infty$. 
 The real part of the permittivity of the silver, gold, 
  and many other metals is well described by 
 the Drude formula in red and infrared spectral range 
 $\Re  \varepsilon_m (\omega) \propto - $
$ (\omega_{p}/\omega)^{2}$
\cite{Johnson1972} and the minimal resonance frequency 
$\omega_{r}^{(m)} \propto $
$ \omega_{p}  (d/a)^{1/4} \ll \omega_{p} $. 

Observation of the plasmon resonance between 
metal cylinders is a difficult experimental problem of 
positioning two nanocylinders parallel each other at 
a nanosize distance. Yet, the investigation of the 
plasmon resonance in the metallic nanocylinder 
placed at nanodistance above a flat metallic surface 
(i.e., $a_{1} = \infty$) is in the center of nowadays 
experimental studies. To obtain the electric field in 
the gap between metallic cylinder and metallic 
surface, we are going to the limit 
$a_{1} \rightarrow \infty$, $d_{1} \simeq
 L^{2}/2 a_{1} \rightarrow 0$. 
The electric field $E_{0}$ is fixed at the surface of the 
large cylinder  ``$a_{1}$" far away from the cylinder  
``$a$" and all the fields are expressed in terms of this 
field. After the limit $a_{1} \rightarrow \infty$ is taken, 
the field $E_{0}$ is the electric field at the metallic 
surface $x =0, \; | y | \gg a $ (see Fig.\,\ref{Fig1}c). Thus, we 
get the electric field in the center coordinates 
$x = y = 0$, i.e., $u = v = 0$. 
The electric field $E(x,y)$ has the maximum at this 
characteristic point just below the cylinder (see 
Fig.\,\ref{Fig1}\,c,d), the jacobian $J$ reduces to the 
scalar $L$, and we obtain by solution of 
Eqs.\,(\ref{AB}) and (\ref{AB1}) and substitution in 
Eq.\,(\ref{F0}):
\vspace{-5pt} 
\begin{equation}\label{E00}
E(0,0) = \lbrace 
{E}_{x0} \left[1 \! +  
\frac{ 8 \varepsilon_{m_{1}}  \Sigma}
{\varepsilon_d - \varepsilon_{m} }
\right]\!, 
{E}_{y0} \left[1 - 
 \frac{8 \varepsilon_{d}  \Sigma}
{\varepsilon_d - \varepsilon_{m} }
\right] 
 \rbrace,
\end{equation}
\vspace{-15pt} 
\begin{equation}\label{SAh}
\Sigma = \sum^{\infty}_{q=1} 
\frac{ (-1)^q q\, h}{(b+1)^q - h}, 
h = \frac{\left(\varepsilon _d-\varepsilon _m\right) 
	\left(\varepsilon _d-\varepsilon_{{m_{1}}}\right)}
{\left(\varepsilon_d+\varepsilon _m\right) 
	\left(\varepsilon_d+\varepsilon _{{m_{1}}}\right)},
\end{equation}
\vspace{-10pt} 
\begin{equation}\label{b}
b = (d + L) (2 a + d + L)/a^{2} \simeq \sqrt{8d/a},
\end{equation}
where $\varepsilon _{m}$, $\varepsilon _{m_{1}}$ 
and $\varepsilon _{d}$ are the permittivities of the 
cylinder, metal plate, and outer space. 
$a$ and $d$ are the cylinder radius and the gap size, 
respectively. The last approximation 
in Eq.\,(\ref{b}) holds for $d \ll a$. 
The system resonates if a denominator in
Eq.\,(\ref{SAh}) almost vanishes, i.e.,
 the dimensionless parameter $\Re h >1 $
and the loss factor 
$\kappa = \Im h  /\vert h \vert  \simeq 2 \varepsilon_d 
\left( \Im \varepsilon_m \left| \varepsilon_m 
\right|^{-2}+ \Im \varepsilon_{m1} 
{\left| \varepsilon_{m1}\right|^{-2}}\right) \ll 1$ 
for 
$\left| \varepsilon_{m} \right|, 
\left| \varepsilon_{m_{1}} \right| \gg 
\varepsilon_{d}$.
 
\begin{figure}[ht!]
\begin{center}
\includegraphics[scale=0.46]{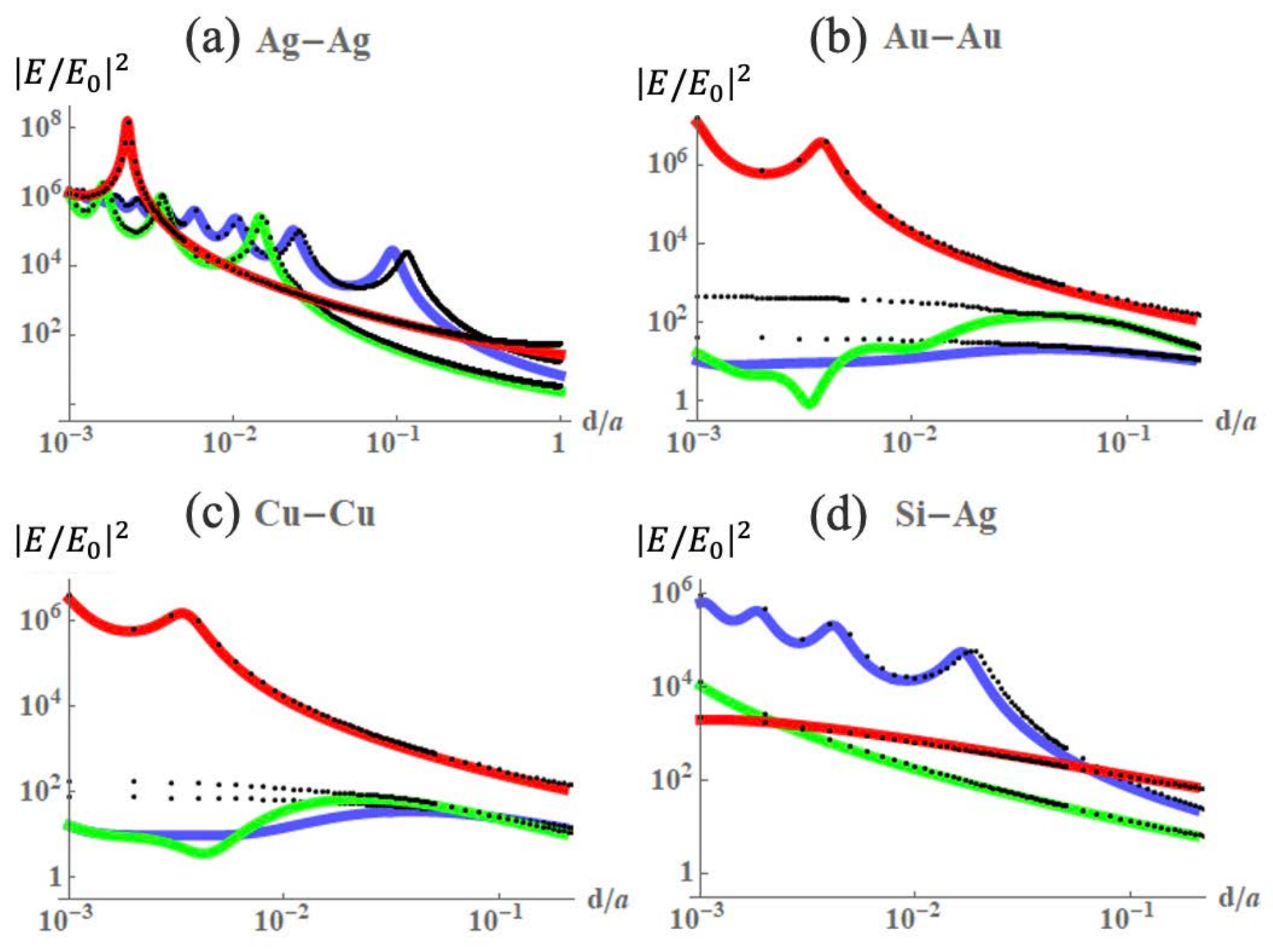}
\caption{Comparison of analytical (lines) and 
numerical (dots) electric field intensities 
${|E/E_{0}|^{2}}$ in the gap between (a) Ag cylinder 
and Ag surface, (b) Au cylinder and Au surface, (c) Cu 
cylinder and Cu surface, (d) Si cylinder and Ag 
surface. Red, green and blue lines correspond to the 
excitation wavelengths of 785 nm/532 nm/405 nm, 
respectively. The excitation beam has an incidence 
angle of $45^{\circ}$, and a $p$-polarization. The 
results are shown in logarithmic scale.
\label{Fig2}}
\end{center}
\end{figure}

In spite of simplicity of Eqs.\,(\ref{E00}) and (\ref{SAh}),
 the gap field has rather rich behavior when the gap vanishes, 
 i.e., $b \rightarrow 0$. Suppose that the parameter $\Re h<1$, 
 then the denominator $n^{(q)} = (b+1)^q - h$ in the sum 
 $\Sigma$ in Eq.\,(\ref{SAh}) is not close to zero for all the values 
 of $q$, and therefore, it can be linearized as 
 $n^{(q)} \simeq  b_{1} q - h_{1}$, 
 where 
 $h_{1} = h - 1$, 
 $ \Re h_{1} < 0 $, $b_{1} =  \log( 1 + b) \simeq b$. 
 Then, $\Sigma$ approximates as 
 $\Sigma \simeq  S(h_{1}, b_{1}) =
  \sum_{q } ^{\infty} \ (-1)^q q/(b_{1} q - h_{1} )$ 
  and can be solved by Laplace transformation and  
  Abel regularization. The function 
  $S$ has simple integral presentation:
\vspace{-5pt} 
\begin{equation}\label{FS1}
\Sigma \simeq	S =
 \frac{h_{1} \left( 
\psi_{1} - \psi_{2} \right)+b_{1}}{2 b^{2}_{1}} 
= -  \int_{0}^{1} \frac{x^{ - \frac{h_1}{b_1}} }{(x+1)^{2}}
 \frac{dx}{b_{1} }
\end{equation}
where 
$  \psi_{1} = 
\psi \left(\frac{b_{1} - h_{1}}{2 b_{1}}\right) $,
$  \psi_{2} = 
\psi\left(-\frac{h_{1}}{2 b_{1}}\right) $ and
$\psi = \Gamma^{\prime}/ \Gamma$ is the 
polygamma function. The sum approximates as 
$\Sigma \simeq$ $-\frac{1}
{4 \left(  b_1 - h_1  \right)}$ 
$\left(\frac{b_1}{2 b_1 -  h_1}+1 \right)$ 
$\rightarrow \frac{1}{ 4 h_{1} }$ 
for the fixed $h_{1}$ and 
$\left| b_{1}/h_{1} \right| \rightarrow 0$. 
That is the electric field tends towards a finite value 
when the gap between cylinder and plate vanishes. 
The integral (\ref{FS1}) converges for 
$\Re h_{1}/b_{1} < 1$. 
In the opposite case, one of the denominators 
$n^{(q)}$ 
in the sum 
$\Sigma$
 in Eq.\,(\ref{SAh}) vanishes for 
 $q =  q_{c} = \log h/\log(1+b)$. 
 We linearize the denominator at this point obtaining 
 $n^{(q)}  \simeq [d \, n^{(q)} /d \, q ](q-q_{c} ) = 
 q b_{2} - h_{2}$, 
 where 
 $b_{2} = h \log (1+b)$ and $h_{2} = h \log h$. 
 The terms in $\Sigma$ with numbers close to 
 $q_{1} =   \Re q_{c} $ 
 make maximum impact to the sum 
 $\Sigma$. 
 The analytical continuation of the function 
 $S(b_{2},  h_{2})$
  to the domain $\Re[h_{2}/b_{2}] <0$ gives :
\vspace{-8 pt} 
\begin{equation}\label{FS2}
\Sigma \simeq -\frac{x^{2}}{b_{2} \, 
	\text{sin}(x)} -S\left(-h_2,b_2\right)
\end{equation}
where 
$x = \pi h_{2} / b_{2} = 
\frac{\pi \log h}{\log(1+b)} \simeq 
\frac{\pi \log \vert h \vert}{\log(1+b)} +
 i \frac{\pi \, \kappa}{\log(1+b)}$. 
 The first resonance term in Eq.\,(\ref{FS2}) 
 determines the singular behavior of the field when 
 the loss factor
$\kappa \ll 1$. Suppose that the 
frequency is fixed, then the sum $\vert \Sigma \vert$ 
oscillates as function of $b \ll \log \left|  h \right|$ 
(see Fig.\,\ref{Fig2}). 

The oscillation period decreases with decreasing of 
the gap size. The local maxima $\vert \Sigma 
\vert_{m}$ are achieved when $\Re x =m \pi$. The 
value of these maxima is given by the following 
expression:
\vspace{-5pt} 
\begin{equation}\label{Smax}
\vert \Sigma \vert_{m}  \simeq 
\frac{	\pi m^{2}}{ \vert h_{2} \vert} 
\sinh\left(\frac{\pi m \kappa}{\log \vert h \vert} 
\right)^{-1}
\end{equation}
This value linearly increases with the number $m$: 
$\vert \Sigma \vert_{m}  \propto m/\kappa$ achieving 
absolute maximum for 
$m = m_{mx} \simeq \log \vert h \vert / \pi \kappa$, 
namely, 
$\vert \Sigma \vert_{mx} \simeq 
\log \vert h \vert / (\pi \kappa^{2} \vert h \vert)$. 
The  oscillations collapse 
with further decrease the gap due to the losses in the 
system and 
$\left|  \Sigma   \right|$ 
exponentially drops down as function of $b$. 
Substituting 
$\vert \Sigma \vert_{mx}$ 
in Eq.\,(\ref{E00}), we obtain the estimation of the 
limit of the electric field 
\vspace{-5pt} 
\begin{equation}\label{Emax}
\left|E_{mx} \right|^{2}  \simeq
 \frac{64   \left| h  \right|^4 \log^{2} | h |   }
{\pi^2    (\Im h)^4} \left|  E_{00}\right|^{2} 
\end{equation}
where $ \left|  E_{00}\right|^{2} 
=\left| E_{0x} \varepsilon_{m1}\right|^2 +
\left| E_{0y} \varepsilon _d\right|^2 $, \!
 $\varepsilon_{m}$, 
$\varepsilon_{m_1}$ and 
$\varepsilon_{d}$ are the permittivities of the metallic 
cylinder, metallic plate and surrounding space, 
respectively. In the red and infrared spectral ranges, 
the metal permittivity is large in absolute value ($ 
\left| \varepsilon_{m} \right|, 
\left| \varepsilon_{m_1} \right| \gg \varepsilon_{d}$, 
$\Im  \varepsilon_{m} / \left| \varepsilon_{m}
 \right|, \Im \varepsilon_{m_1}  
 \left| \varepsilon_{m_1} \right| \ll 1$) 
 and Eq.\,(\ref{Emax}) simplifies:
\vspace{-1 pt} 
\begin{equation}\label{Emax1}
\left| E_{mx}^{2} \right|  \propto 
\frac{ \left| \varepsilon_{m}^{6} \right| 
\left| \varepsilon_{m_1}^{4} \right|
\left(\left| \varepsilon_m \right| + 
\left| \varepsilon_{m_1} \right| 
\right)^2 }
{\varepsilon_{d}^{2}   \left(\left| 
\varepsilon_{m}^{2}\right|  \Im \varepsilon_{m_1} + 
\left|\varepsilon_{m_1}^{2} \right| \Im  \varepsilon_m  
\right)^{4}}  
 \left|  E_{00} \right|^{2} 
\end{equation}
The maximum field enhancement estimates as 
$\left|E_{mx} \right| \propto 
\left| \varepsilon_m \right|^{3} 
\varepsilon_{d}^{-1} \left( \Im \varepsilon_{m}\right)
 ^{-2} \left| E_{0} \right|$ 
 in the gap between cylinder and plate 
 made of the same metal. For instance, 
 the field could be as large as 
 $\left|E_{mx} \right|  > 10^{3} \left| E_{0} \right|$
  in the gap between the silver plate and a silver 
  cylinder for the green light $\lambda = 532\,nm$ 
  (Fig.\,\ref{Fig2}a). It is an upper limit since the field could be 
  restricted by radiation loss and
   the spatial dispersion of the electric charge
  in subnanometer gap \cite{Ciraci2012}.
\section{Analytical theory vs. Simulations}
In any real experiment, a layer of investigated
molecule is placed in the gaps formed by metallic
particles distributed over a metallic mirror. 
We use full-scale computer 
simulations in COMSOL environment  to
simulate the periodic array 
of metallic nanocylinders with the radius 
$a = 10\,nm$,  period of $D=200\,nm$,
and for ``large" metallic nanocylinders,
$a = 50\,nm$ and $D=200\,nm$.
The cylinders are placed over the metal surface and are
excited by \textit{p}-polarized light
with an incidence angle of $45^{\circ}$ 
similar to \cite{MubeenS2012}. 
The periodic boundary conditions and periodic ports
are used to compute the reflectance as well as 
 local electric field.
Thus obtained, the enhancement of the gap electric field is
in agreement with analytical theory as shown in 
Figs.\ref{Fig2}\,a-c. The gap field could also achieve 
large values in the metal-semiconductor (dielectric) 
gap (Fig.\ref{Fig2}d).
\begin{figure}[ht!]
\begin{center}
\includegraphics[scale=0.36]{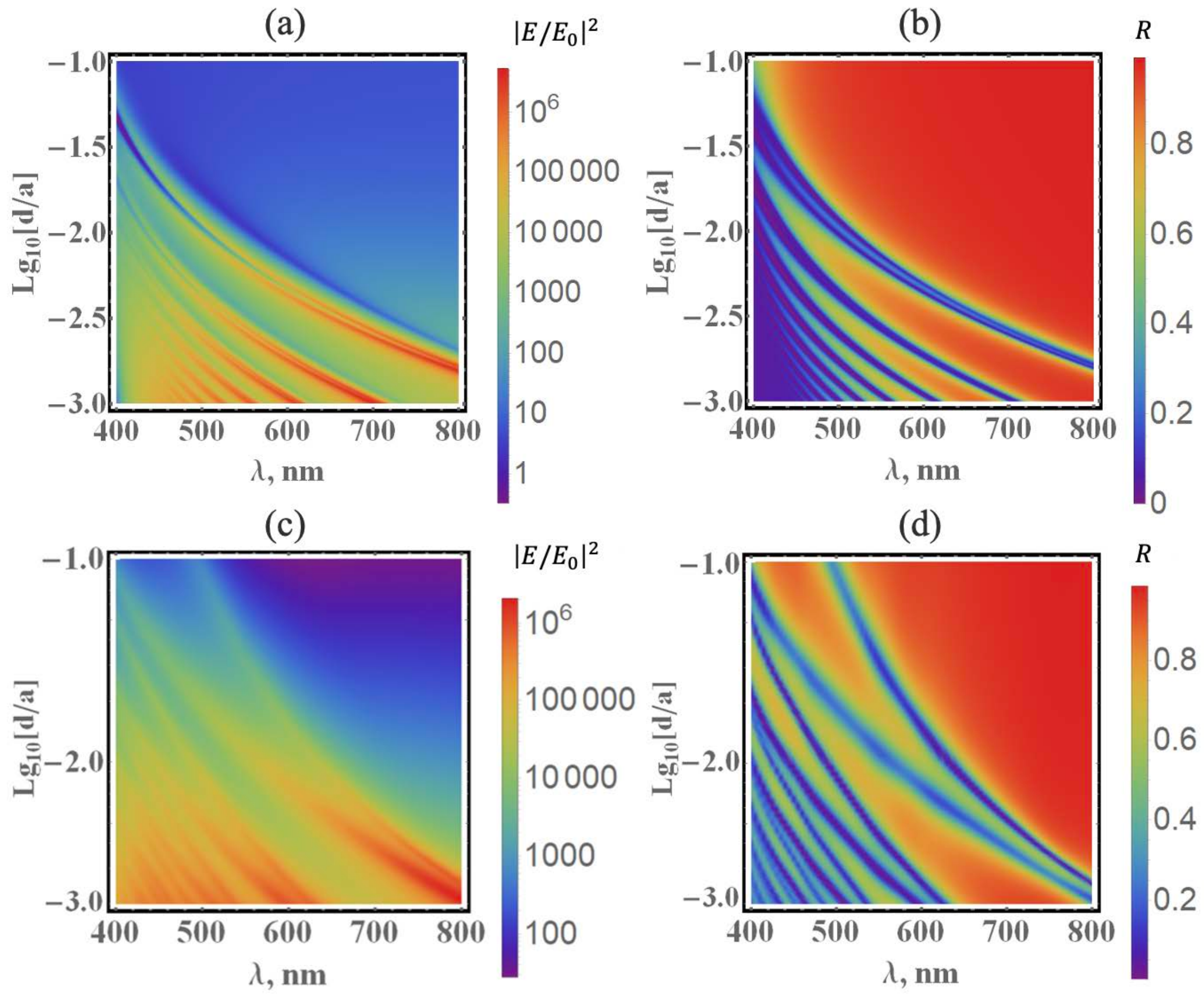}
\caption{Analytical results in the gap between Ag cylinder 
and Ag surface as function of wavelength and $d/a$ ratio: 
(a) electric field intensity ${|E/E_{0}|^{2}}$ 
and (b) reflectance. Numerical results for ``large" 
silver cylinder with a radius of $a = 50\,nm$ and a
period of $D =  200\,nm$: (c) electric field intensity 
${|E/E_{0}|^{2}}$ and (d) reflectance. 
\label{Fig3}}
\end{center}
\end{figure}
The field enhancement  in the gap between a cylinder 
and the surface could revile itself in the 
anomalous reflectance (Fig.\,\ref{Fig3}\,b,d).

To compare the theory with computer simulations, we fix the 
electric field $E_{s}$ at the point 
$x= d+2 a, y=0$ on the surface of 
a cylinder and calculate the field 
$E (x,y)$ in all the space by using 
our analytical theory. To simplify the consideration, the direct 
interaction between cylinders is neglected 
since the electric field is 
localized in the nanogaps with size
$ d \ll a $. Then, we average the field 
over the layer $0< x < d+2 a$ and introduce the effective 
permittivities 
$\varepsilon^{(e)}_{xx}  =
 \left\langle \varepsilon(x,y) E_{x} (x, y) \right\rangle / 
 \left\langle E_{x} (x, y)\right\rangle$ 
 and 
$\varepsilon^{(e)}_{yy}  =
\left\langle \varepsilon(x,y) E_{y} (x, y) \right\rangle / 
\left\langle E_{y} (x, y) \right\rangle$. 
Electromagnetic wave with an amplitude 
$E_{in}$ and the  $p$-polarization is incident on the layer placed 
above the metal plate. Maxwell equations are solved in terms of 
the amplitude $E_{in}$ in order to find electromagnetic field in the 
layer with thickness $d+2 a$ and permittivity 
$\hat{\varepsilon}^{(e)}$. In particular, we find the electric field 
$E_{0} =\{E_{x 0}, E_{y 0} \}$
at the interface between the layer and the metal plate (see Fig.\,\ref{Fig1}c). 
In the dilute case $a \ll D \ll \lambda$, the field $ E_{0} $
equals to the field that were on the metal surface without 
cylinder, namely 
$ E_{0} = E_{in}
\{\frac{\varepsilon_{m_1} \sin 2 \theta}
{\sqrt{\varepsilon_d \varepsilon_{m_1}
		-\varepsilon_{d}^{2} \sin^2 \theta}+
	\varepsilon_{m_1}\cos \theta },
\frac{2 \cos \theta \sqrt{\varepsilon_d \varepsilon_{m_1}-
		\varepsilon_{d}^{2} \sin^2 \theta }}
{\sqrt{\varepsilon_d \varepsilon_{m_1}-
		\varepsilon_{d}^{2} \sin^2 \theta}+\varepsilon_{m_1} \cos \theta}\}$,
where $ \theta $ is the angle of  the incidence.
This field is introduced in Eq.\,(\ref{E00}) 
in order to obtain the analytical 
esteem of the electric field in the nanogap. 
The theoretical results are in accordance with 
full-scale computer simulations for small cylinders ($k a \ll 1$). 
The results of our quasistatic theory being extrapolated
to  $ k a \sim  1 $ are still in a qualitative agreement with
 computer simulations of ``large" nanocylinders (Fig.\ref{Fig3}).
 The agreement is better for the smaller gap size
 $ d $ and higher plasmon resonance $ q $.
 The theory holds when the plasmon size 
 $ L_{q} \sim \sqrt{ad}/q $ is smaller than the skin  depth
 $ L_{q}  \ll \frac{1}{k \sqrt{\left| \varepsilon_{m}  \right| } } $. 
Figures \ref{Fig2} and \ref{Fig3} show that 
the largest field enhancement is achieved in red and infrared
spectral bands, where optical loss in metal is relatively small.
On the other hand, the silver-silicon system resonates for 
$ \lambda < 550 \,nm $ when the parameter 
$\Re h $ in Eq.\,(\ref{SAh}) becomes larger than one.

\vskip 0.1 cm
In summary, we present a theory of the plasmons 
excited in the gap between a metal or dielectric nanocylinder 
and a metallic or dielectric plate. The resonance conditions for the 
gap plasmon are found as function of the metal permittivity and 
the gap size. When the cylinder-on-mirror cavity is excited by 
the incident light, the gap electric field increases and 
oscillates with decreasing the gap size achieving its limit value.
Excitation of the gap plasmons results in the dips 
in the reflectance. The ``$ q $-th" minimum of the reflectance
 corresponds to the excitation of  the ``$ q $-th" order plasmon.
 Fluctuations in the reflectance were probably  observed, e.g.,
 in experiments \cite{Ciraci2012,KanipeK2017}.
The quasistatic theory qualitatively describes 
the field enhancement up to $k a  \leq 1$, since the 
field concentrates in the nanogap whose size is much 
smaller than $\lambda$. 
Finally, this analytical theory can be used to 
design new SERS substrates and other optical sensors.

\vskip 0.1 cm
The work is supported by Russia Fund Basic Research
 grant 20-21-00080.


\providecommand{\latin}[1]{#1}
\makeatletter
\providecommand{\doi}
  {\begingroup\let\do\@makeother\dospecials
  \catcode`\{=1 \catcode`\}=2 \doi@aux}
\providecommand{\doi@aux}[1]{\endgroup\texttt{#1}}
\makeatother
\providecommand*\mcitethebibliography{\thebibliography}
\csname @ifundefined\endcsname{endmcitethebibliography}
  {\let\endmcitethebibliography\endthebibliography}{}
\begin{mcitethebibliography}{0}
\providecommand*\natexlab[1]{#1}
\providecommand*\mciteSetBstSublistMode[1]{}
\providecommand*\mciteSetBstMaxWidthForm[2]{}
\providecommand*\mciteBstWouldAddEndPuncttrue
  {\def\EndOfBibitem{\unskip.}}
\providecommand*\mciteBstWouldAddEndPunctfalse
  {\let\EndOfBibitem\relax}
\providecommand*\mciteSetBstMidEndSepPunct[3]{}
\providecommand*\mciteSetBstSublistLabelBeginEnd[3]{}
\providecommand*\EndOfBibitem{}
\mciteSetBstSublistMode{f}
\mciteSetBstMaxWidthForm{subitem}{(\alph{mcitesubitemcount})}
\mciteSetBstSublistLabelBeginEnd
  {\mcitemaxwidthsubitemform\space}
  {\relax}
  {\relax}

\end{mcitethebibliography}


\begin{thebibliography}{50}
\bibitem{Vidal1996}
F. J. Garcia-Vidal, and J. Pendry, Phys. Rev. Lett. 
\textbf{77}, 1163 (1996).
\bibitem{Gresillon1999}
S. Gr\'esillon \textit{et al.}, Phys. Rev. Lett. 
\textbf{82}, 4520 (1999).
\bibitem{Ciraci2012}
C. Ciraci \textit{et al.}, Science \textbf{337}, 
1072 (2012).
\bibitem{Moreau2012} 
A. Moreau \textit{et al.}, Nature \textbf{492}, 86 (2012).
\bibitem{Echtermeyer2011} 
T. J. Echtermeyer \textit{et al.}, Nat. Commun. 
\textbf{2}, 458 (2011).
\bibitem{ZhangY2011} 
Y. Zhang \textit{et al.}, Nano Lett. \textbf{11}, 
5519 (2011).
\bibitem{Savage2012}
K. J. Savage \textit{et al.}, Nature 
\textbf{491}, 574 (2012).
\bibitem{MertensJ2013}
J. Mertens \textit{et al.}, Nano Lett. 
\textbf{13}, 5033 (2013).
\bibitem{CaiH2016}
H. Cai \textit{et al.}, Opt. Express 
\textbf{24}, 20808 (2016).
\bibitem{MubeenS2012}
S. Mubeen et al., Nano Lett. \textbf{12}, 2088 (2012)
\bibitem{ChikkaraddyR2017} 
R. Chikkaraddy \textit{et al.}, ACS Photonics 
\textbf{4}, 4649 (2017).
\bibitem{ChikkaraddyR2018} 
R. Chikkaraddy \textit{et al.}, Nano Lett. \textbf{18}, 
405 (2018). 
\bibitem{DemetriadouA2017}
A. Demetriadou \textit{et al.}, ACS Photonics 
\textbf{4}, 2410 (2017).
\bibitem{SealK2005} 
K. Seal \textit{et al.}, Phys. Rev. Lett. \textbf{94}, 
226101 (2005).
\bibitem{Genov2004}
D. Genov,  \textit{et al.},
Nano Lett. \textbf{4}, 153 (2004).
\bibitem{IvanovA2012}
A. Ivanov \textit{et al.}, Appl. Phys. A 
\textbf{107}, 17 (2012).
\bibitem{Frumin2013}
L. L. Frumin, A. V. Nemykin, S. V. Perminov, 
and D. A. Shapiro, J. Opt. \textbf{30}, 2048 (2013).
\bibitem{LiuZ2013}
Z.-q. Liu \textit{et al.}, Plasmonics \textbf{8}, 
1285 (2013).
\bibitem{LiuG2014}
G.-q. Liu \textit{et al.}, Phys. Chem. Chem. Phys. 
\textbf{16}, 4320 (2014).
\bibitem{Liu2014}
G.-q. Liu \textit{et al.}, Opt. Commun. 
\textbf{316}, 111 (2014).
\bibitem{Rasskazov2013}
I. L. Rasskazov, V. A. Markel, and S. V. Karpov, 
Opt. Spectrosc. \textbf{115}, 666 (2013).
\bibitem{DingF2018}
F. Ding, Y. Yang, R. A. Deshpande, 
and S. I. Bozhevolnyi, Nanophotonics 
\textbf{7}, 1129 (2018).
\bibitem{DongL2017}
L. Dong \textit{et al.}, Nano Lett. \textbf{17}, 
5768 (2017).
\bibitem{Kneipp2017}
J. Kneipp, ACS Nano \textbf{11}, 1136 (2017).
\bibitem{HuY2017}
Y. Hu \textit{et al.}, ACS Nano \textbf{11}, 5558 (2017).
\bibitem{AndreouC2016}
C. Andreou \textit{et al.},  ACS Nano 
\textbf{10}, 5015 (2016).
\bibitem{NechaevaNL2020}
N. L. Nechaeva \textit{et al.}, Anal. Chim. Acta 
\textbf{1100}, 2506 (2020).
\bibitem{ChonH2014}
H. Chon \textit{et al.}, Chem. Commun. 
\textbf{50}, 1058 (2014).
\bibitem{FuQ2015}
Q. Fu \textit{et al.}, ACS Appl. Mater. Interfaces
 \textbf{7}, 13322 (2015).
\bibitem{JiangT2018} 
T. Jiang \textit{et al.}, J. Am. Chem. Soc. 
\textbf{140}, 15560 (2018).
\bibitem{LiuG2018} 
G. Liu \textit{et al.}, Sol. Energy Mater. Sol. Cells
 \textbf{186}, 142 (2018).
\bibitem{LuX2018} 
X. Lu \textit{et al.}, Chem. Mater. 
\textbf{30}, 1989 (2018).
\bibitem{MaC2016} 
C. Ma, Q. Gao, W. Hong, J. Fan, and J. Fang, 
Adv. Funct. Mater. \textbf{27}, 1 (2016).
\bibitem{NamJM2016} 
J.-M. Nam, J.-W. Oh, H. Lee, and Y. D. Suh, 
Acc. Chem. Res. \textbf{49}, 2746 (2016).
\bibitem{KanipeK2017} 
K.N. Kanipe et al. 
J. Phys. Chem. C \textbf{121} 14269 (2017)
\bibitem{PanR2018} 
R. Pan \textit{et al.}, Nanoscale \textbf{10}, 
3171 (2018).
\bibitem{ShinY2015} 
Y. Shin, J. Song, D. Kim, and T. Kang, Adv. Mater. 
\textbf{27}, 4344 (2015).
\bibitem{SigleDO2015} 
D. O. Sigle \textit{et al.}, ACS Nano \textbf{9}, 825 
(2015).
\bibitem{YooD2018} 
D. Yoo \textit{et al.}, Nano Lett. \textbf{18}, 1930 
(2018).
\bibitem{ZhouJ2016} 
J. Zhou \textit{et al.}, ACS Nano \textbf{10}, 
11066 (2016).
\bibitem{KraftM2015} 
M. Kraft \textit{et al.}, Phys. 
Rev. X \textbf{5}, 031029 (2015).
\bibitem{Johnson1972} 
P. B. Johnson, and R. W. Christy, Phys. Rev. B 
\textbf{6}, 4370 (1972).
\end{thebibliography}
\end{document}